# Activation Products from Copper and Steel Samples Exposed to Showers Produced by 8 GeV Protons Lost in the Fermilab Main Injector Collimation System

**Bruce C. Brown, Nikolai V. Mokhov and Vitaly S. Pronskikh**
Fermi National Accelerator Laboratory
Batavia, IL, USA

**Abstract**

*In conjunction with efforts to predict residual radiation levels in the Fermilab Main Injector, measurements of residual radiation were correlated with the time history of losses. Detailed examination suggested that the list of radioactive isotopes used for fitting was incomplete. We will report on activation studies of magnet steel and copper samples which we irradiated adjacent to the Fermilab Main Injector collimation system. Our results identified several additional radioactive isotopes of interest. The MARS15 studies using a simplified model are compared with measurements. The long half-life isotopes will grow in importance as operation stretches to a second decade and as loss rates rise. These studies allow us to predict limits on these concerns.*

**Introduction**

In response to a demand for high intensity operation of the Fermilab Main Injector to provide protons for producing neutrinos, an intense effort [1] to control activation due to beam loss has been carried out. While monitoring the activation and cool down to allow planning for accelerator repairs and upgrades, we found that crude models [2] which acknowledge contributions from three radioactive isotopes of manganese were inadequate to explain some detailed measurements [3]. Using an available HPGe spectrometer, we have measured the isotopes produced in a copper sample and a sample of the magnet lamination steel when activated near the collimation system. We have constructed a simplified model in MARS [4] to explore the activation in two locations with two different momentum spectra.

**Figure 1: Simple Fit to Residual Radiation**

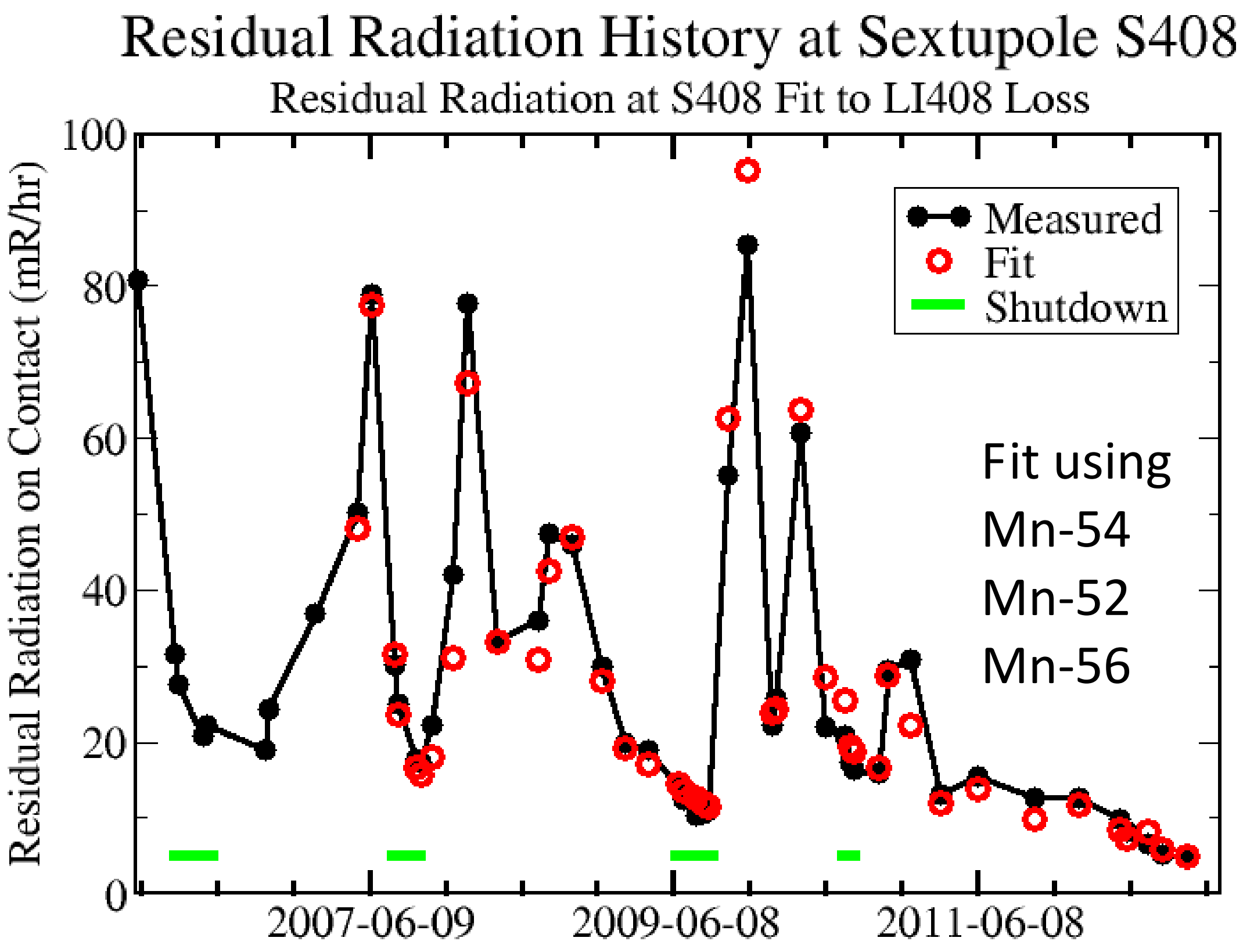

**Studies of Activation History**

Using measurements of residual radiation at a series of 126 (later 142) locations identified by bar coded tags, a simple fitting procedure provided a formula for radiation cool down sufficient for planning work in the accelerator enclosure. The recorded beam loss at a nearby beam loss monitor, weighted by the half-life of the isotope considered, is used in a linear model to fit the residual radiation measurements. A typical result is shown in Figure 1.

**Figure 2: Secondary Collimator Showing Monitoring Locations**

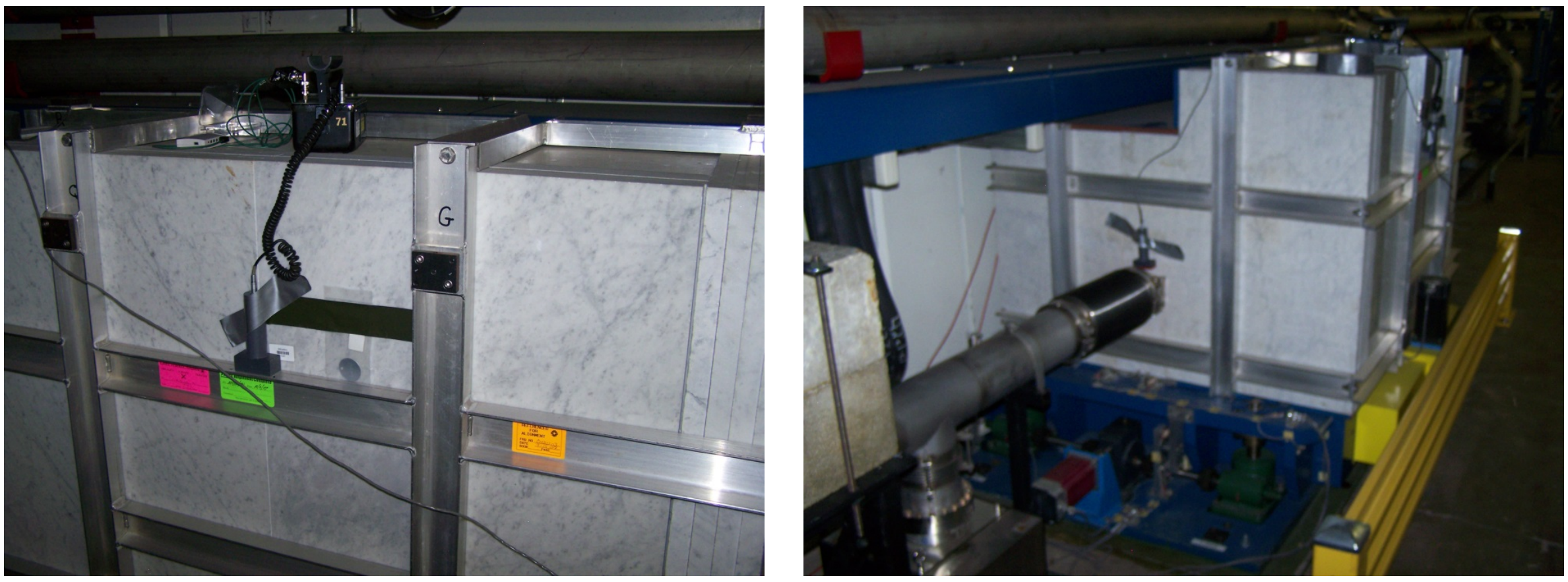

### *Activation Measurements with Copper and Steel Tags*

During machine down time, cool down studies using Geiger counters (see Figure 2) measured the residual radiation [3] for times from one to many days. As more detailed studies showed that the fit required half-life values intermediate between Mn-52 (5.59 day) and Mn-54 (312 day), we initiated a study of activation employing an available high purity Germanium (HPGe) spectrometer [5]. This system had been employed for monitoring using Al Tags (as seen in Figure 2) [6]. 1.5 inch tags of pure Cu and of steel from the magnet laminations were fabricated. These were placed on the side (left photo of Figure 2) and at the downstream end (right photo of Figure 2) of the third secondary collimator in the MI300 straight section [7]. Lost proton beam which was captured by this collimator produced a spectrum of electromagnetic and hadronic secondaries which activates materials in the area. Samples were placed and removed during available access times and delivered to the Fermilab Radiation Analysis Facility (RAF) for measurement. Activation times were varied and smaller samples prepared to allow measurements with a limit of 1 mR/hr at the HPGe spectrometer.

### ***Decay Correction for Activation Measurements***

Analysis of activation measurements usually rely on the 'activation formula' [8] (Equation 1) which assumes a uniform activation rate. MARS calculations assume a uniform activation so for a specific isotope, one can apply the activation formula to translate the results to the desired activation and cool down time.

$$S_A(t_c)(pCi/gm) = \frac{N_A \sigma_I}{3.7\times 10^{-2} A_T} \frac{d\phi}{dt} (1 - e^{-\frac{t_i}{\tau_I}}) e^{-\frac{t_c}{\tau_I}} \qquad (1)$$

where $S_A$ is the activation of isotope I, $N_A$ is Avogadro's number, $A_T$ is the nucleon number for the isotope I, $\sigma_I$ is the cross section for production of isotope I and $\phi$ is the fluence of activating particles, $\tau_I$ is the lifetime (corresponding with the half-life $t_{1/2}$), $t_i$ is the time of activation and $t_c$ the time allowed for cool down. The measurements were corrected for decay during activation using reading from a nearby Beam Loss Monitor (BLM) which was recorded every Main Injector cycle (typically 2.2 seconds) which allowed accounting for the details of beam intensity and beam loss variations [7]. Activation was reported in the 'instantaneous' activation limit for which $\phi$

= $t_i$ $d\phi/dt$ in the limit $t_i \rightarrow 0$. Measurement results from RAF are corrected to receipt of the sample ($t_c$=0). Equilibrium activation is of interest for long term exposure. In this case, one assumes ($t_c$=0) and takes the limit $t_i$ >> $t_{1/2}$.

**Figure 3: Toy Model for MARS simulation (Beam from Left)**

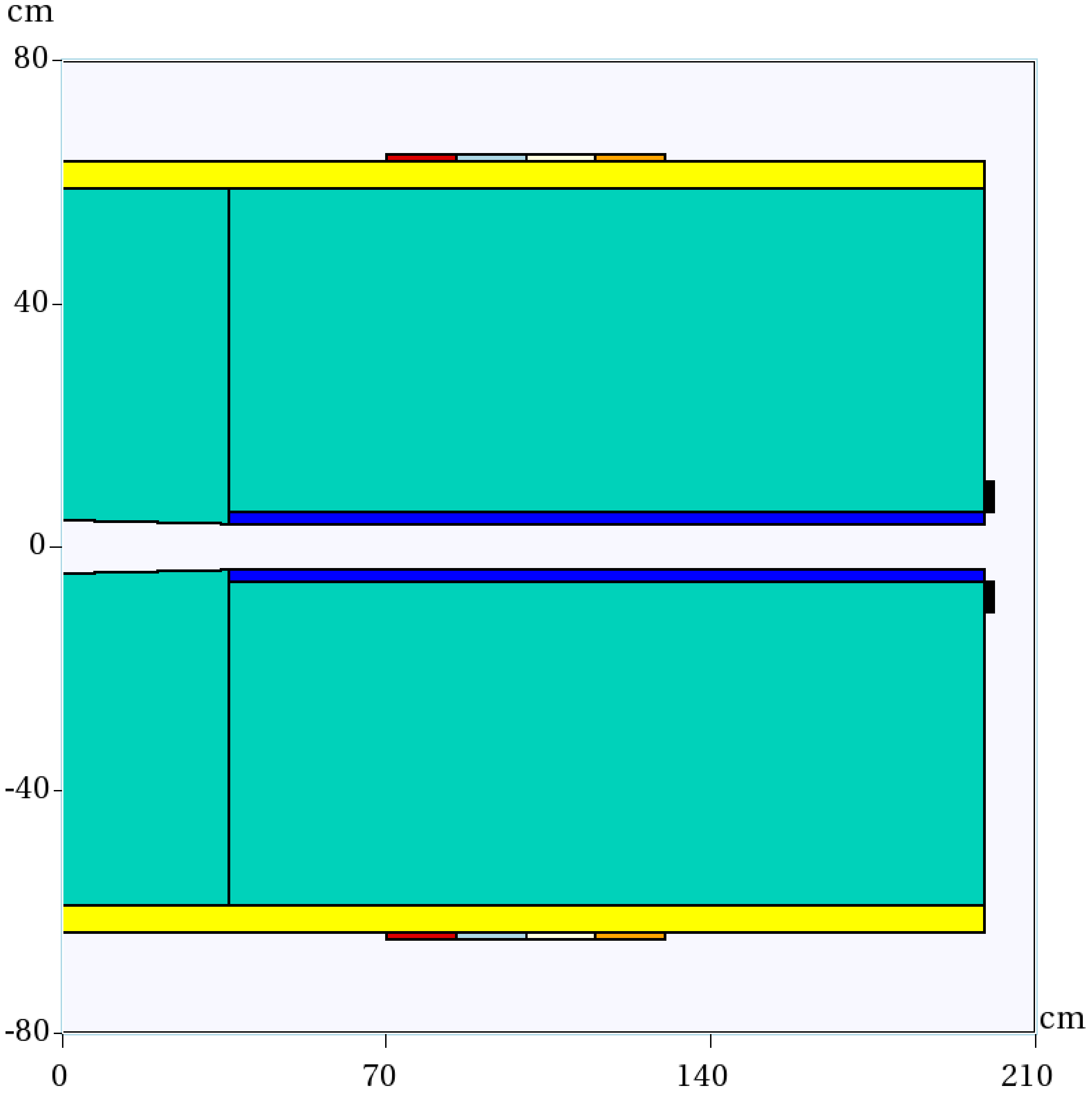

## MARS15 Model and Collimator Details

A complete model of the collimation system has been constructed in MARS [4] but for this study we employ a simplified geometry ('toy model') which supports the essential features while requiring much less computation. The toy collimator model (Figure 3) employs a ~200 cm long cylinder with a 3.8 cm inner bore radius and a 63.5-cm outer radius. The first 36 cm of the inner bore are tapered from 4.4 cm at the entrance to 3.8 cm. The collimator main body in the model is made of the yoke steel (green), with a stainless steel (blue) inner and a marble (yellow) outer layer. The $1.25\times10^{12}$ p/s proton beam strikes the inner surface of the bore at the end of the tapered part (30 cm from the entrance to the collimator). The 'shielded' samples are placed onto the outer surface of the marble layer, and the 'unshielded' ones near the exit at the downstream end of the collimator. All the samples are cylinders.

The two sample locations provide different spectra for the activating secondaries. The tapered upstream in both model and real collimator is intended to intercept all losses in a short longitudinal region. The side (shielded) location is similar for the model and measurements. The downstream (unshielded) location in the collimator is activated by the losses at the end of the tapered region but also by small losses near the downstream end of the collimator. Details of this are under investigation. As a result, the toy model will provide a more energetic (harder) spectrum of secondaries but is not intended as a careful representation of the measurement configuration.

**Figure 4a: Gamma Spectra from HPGe Spectrometer**

**490 to 820 keV Spectrum**

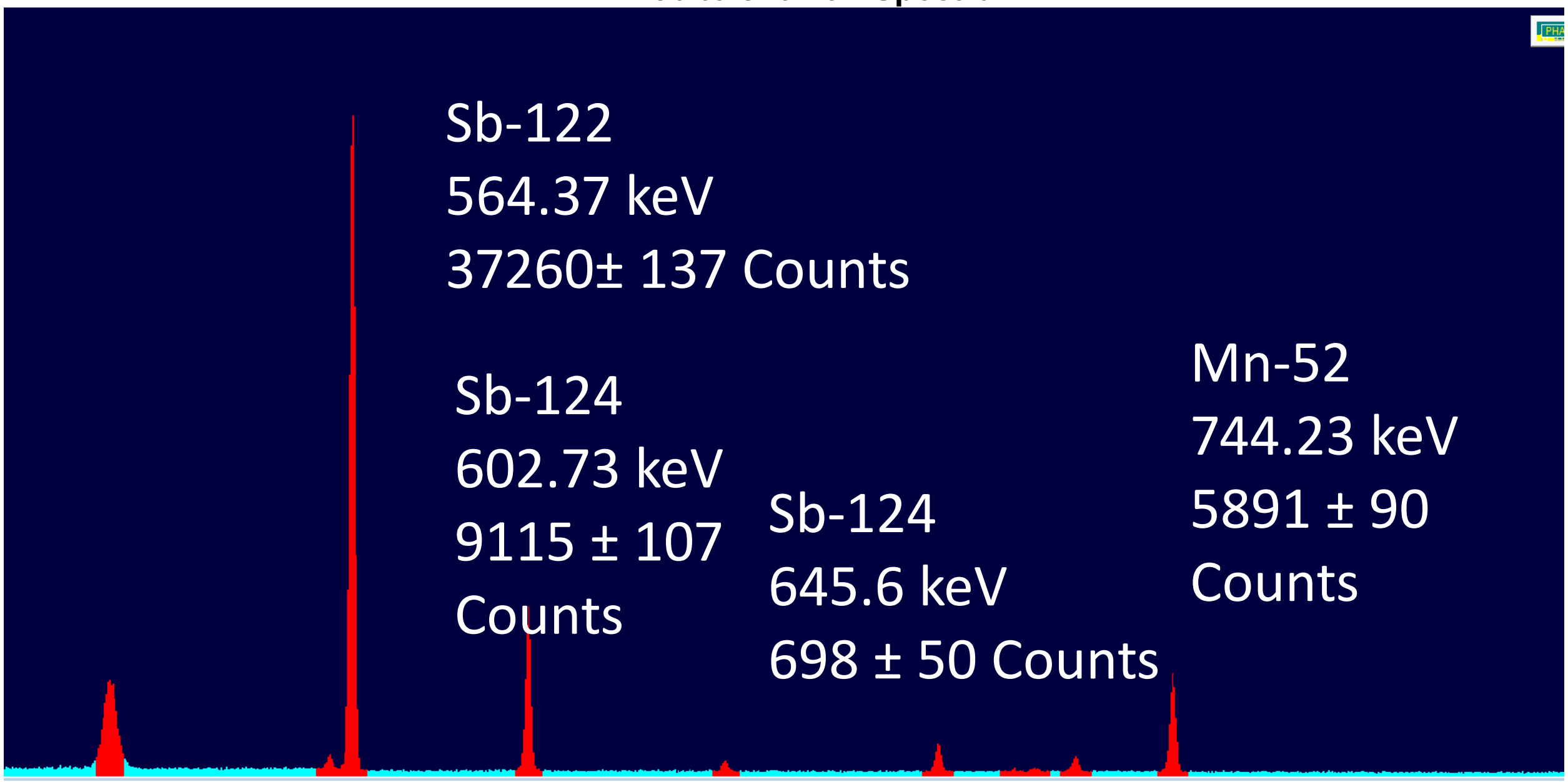

**Figure 4b: Gamma Spectra from HPGe Spectrometer**

**1000 to 1325 keV Spectrum**

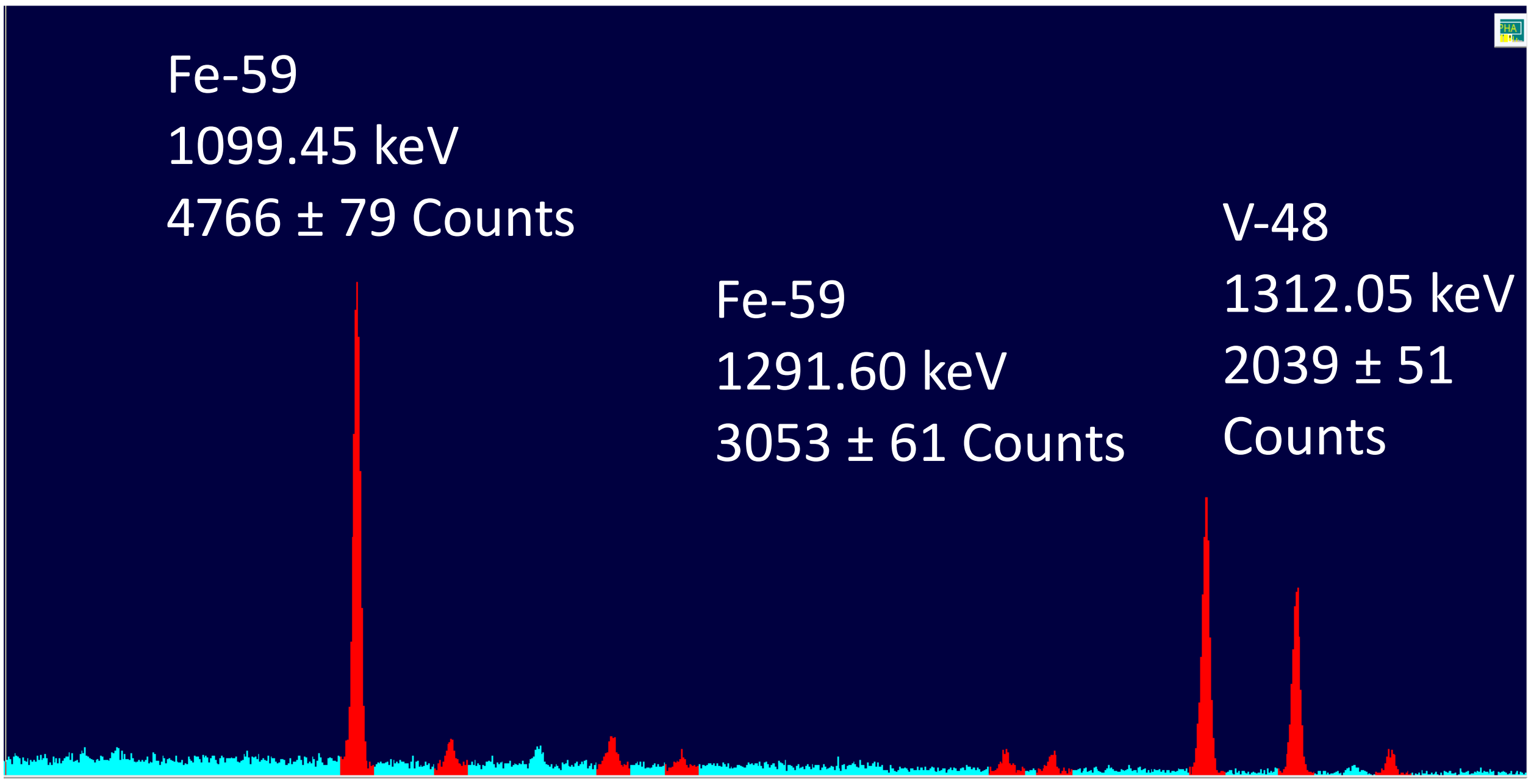

### Results of Measurement and Simulation

To explore the distribution of isotopes produced in this activation study, we recognize that the normalization and spectra are imperfectly reproduced in the 'toy model' MARS simulation. We choose to normalize to dominant isotopes for each sample using Cu-64 for the copper sample and Mn-54 for the Fe and yoke steel samples. The Mn-54 dominates the activation after a few days and for the 'shielded' location, the measurements of three samples agree to 1.2% when normalized to the half-life weighted BLM reading. Figure 4 shows a part of the spectra for one steel sample. Using the analysis software, the results for each measured isotope were obtained using multiple gamma lines when available. These were combined using known branching fractions and corrected for decay from the sample removal time. Using these results and the half-life weighted BLM reading, results are tabulated for the ratio of activation over BLM reading corrected to the instantaneous exposure limit [7]. These values were then corrected to the 30 day exposure with 2 hour decay condition to match the choice used in the MARS calculations. Measurements and MARS calculation results (shielded or unshielded exposures separately) were normalized to Cu-64 (copper samples) or Mn-54 (steel). Results are shown in the figures below.

We note that the Main Injector yoke lamination steel had a 0.3% by weight of Antimony (Sb). Activation of this to Sb-122 and Sb-124 was measured. This was not included in the MARS simulations. Further study of this will be required.

## Conclusions

We note that the relative abundance of radioactive isotopes produced in this secondary flux is not precisely predicted by the current toy model MARS simulation. The measurements have identified isotopes which contribute to the observed activation. We find that using the Fe-59 or Cr-51 half-life for the weighted fit contributes to the shape in ways which will not impact planning for tunnel work radiation limits.

**Acknowledgements**

Fermi National Accelerator Laboratory is operated by Fermi Research Alliance, LLC under Contract No. DE-AC02-07CH11359 with the United States Department of Energy.

We thank Vernon Cupps, Meka Francis, Gary Lauten and Matt Quinn for assistance in exposing the samples, measuring the activation and aiding in presentation of the results.

**Figure 5: Steel Activation Normalized to Mn-54 with 30 Day exposure, 2 hour cool down**

**Upper Figure for Shielded Location, Lower Figure for Unshielded Location**

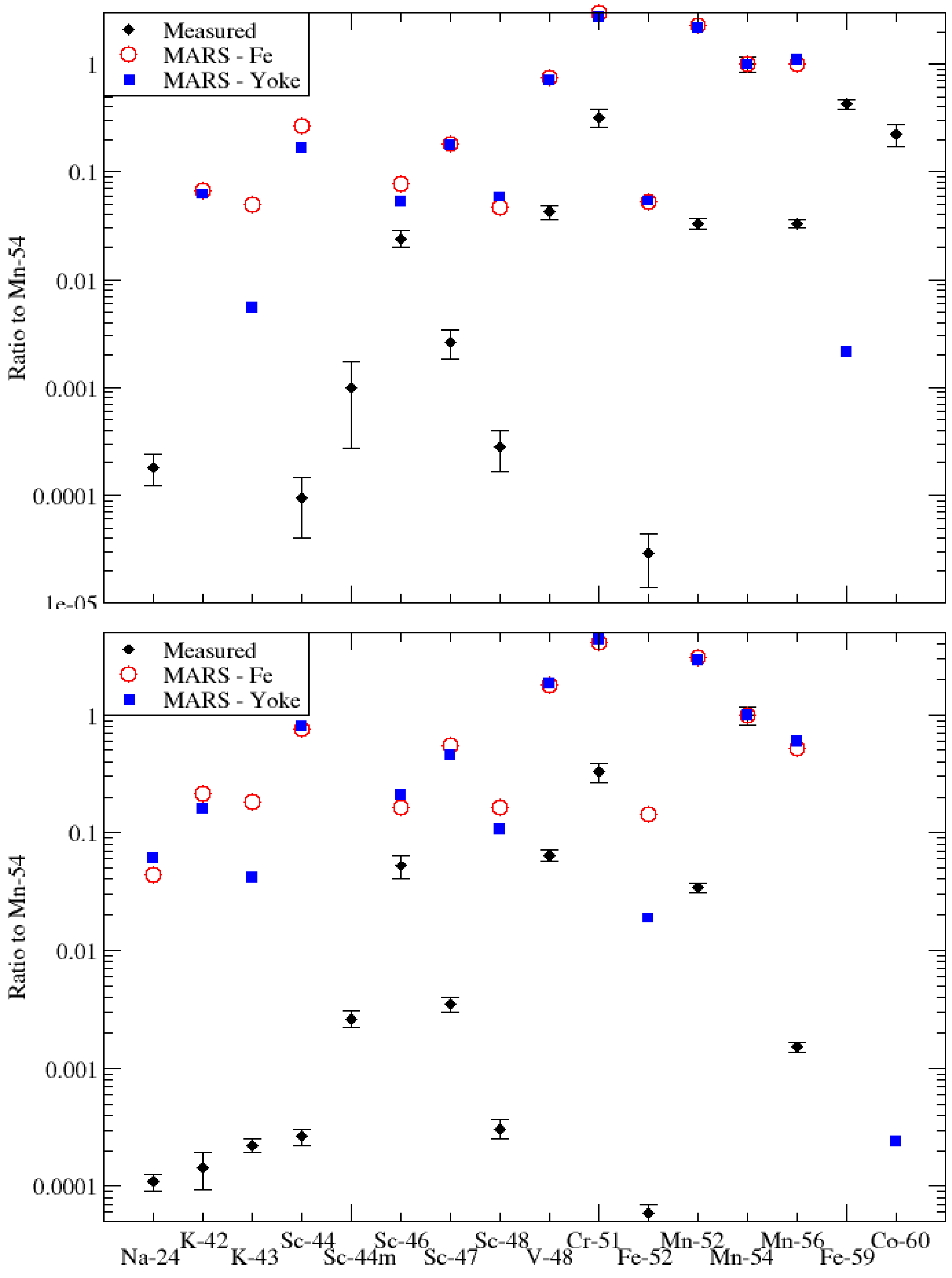

**Figure 6: Copper Activation Normalized to Cu-64 with 30 Day exposure, 2 hour cool down**

**Upper Figure for Shielded Location, Lower Figure for Unshielded Location**

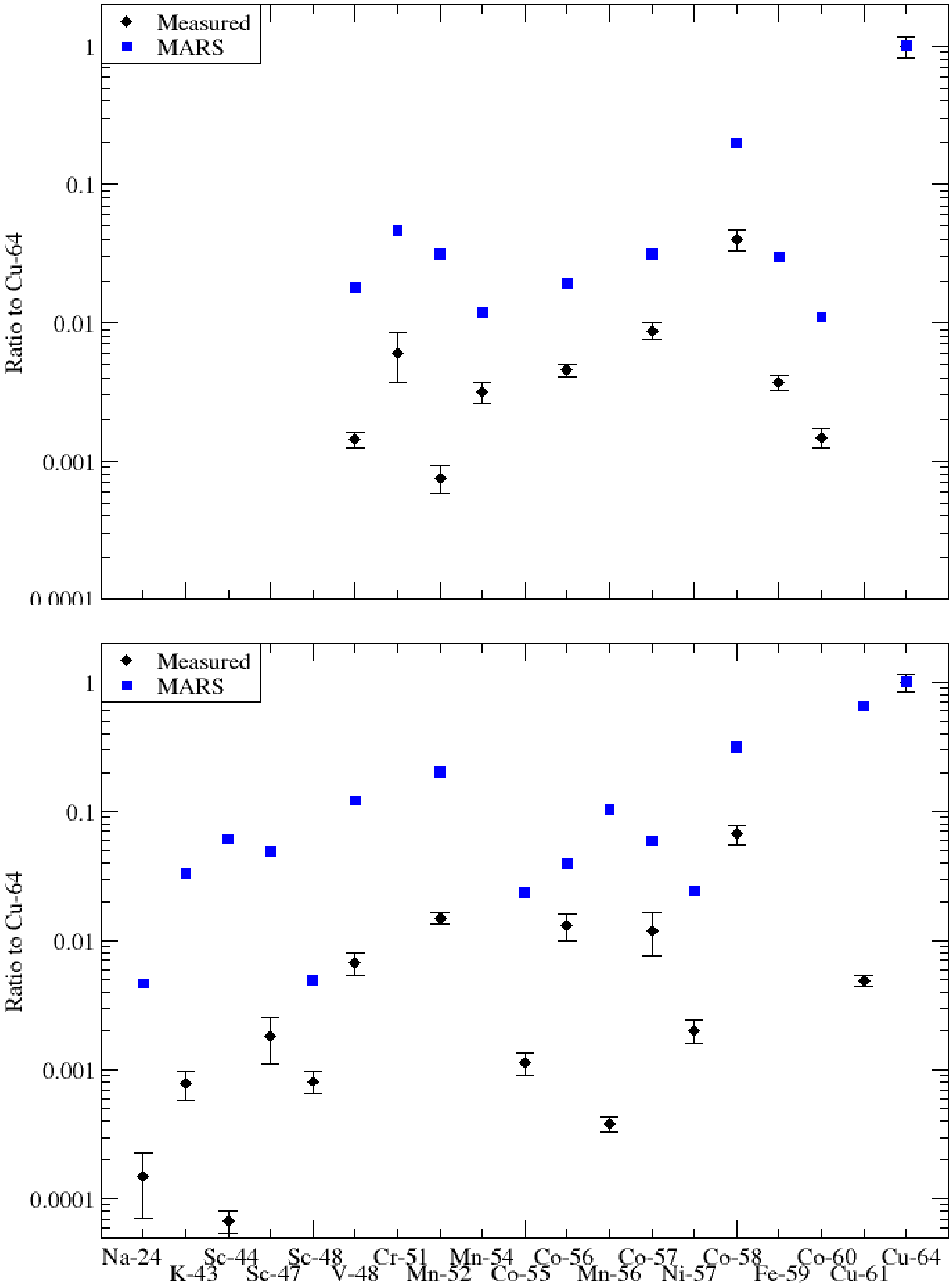